\documentclass[preprint, prd, aps, showpacs]{revtex4-1}

\usepackage{graphicx}
\usepackage{amsmath,amssymb}
\usepackage[usenames,dvipsnames]{color}
\usepackage[colorlinks=true, citecolor=blue, linkcolor=WildStrawberry]{hyperref}
\usepackage{epstopdf}

\newcommand{\drm}{{\rm d}}
\newcommand{\irm}{{\rm i}}
\newcommand{\beq}{\begin{equation}}
\newcommand{\eeq}{\end{equation}}
\newcommand{\bdm}{\begin{displaymath}}
\newcommand{\edm}{\end{displaymath}}

\DeclareFontFamily{OT1}{pzc}{}
\DeclareFontShape{OT1}{pzc}{m}{it}{<-> s * [1.10] pzcmi7t}{}
\DeclareMathAlphabet{\mathpzc}{OT1}{pzc}{m}{it}

\graphicspath{{./plots/}}
\begin{document}

\title{Constraining the gravitational wave energy density of the Universe in the Range 0.1\,Hz to 1\,Hz using the Apollo Seismic Array}

\author{Michael Coughlin}
\affiliation{Department of Physics, Harvard University, Cambridge, MA 02138, USA}

\author{Jan Harms}
\affiliation{INFN, Sezione di Firenze, Sesto Fiorentino, 50019, Italy}


\begin{abstract}

In this paper, we describe an analysis of Apollo era lunar seismic data that places an upper limit on an isotropic stochastic gravitational-wave background integrated over a year in the frequency range 0.1\,Hz -- 1\,Hz. We find that because the Moon's ambient noise background is much quieter than that of the Earth, significant improvements over an Earth based analysis were made. We find an upper limit of $\Omega_{\rm GW}<1.2\times 10^{5}$, which is three orders of magnitude smaller than a similar analysis of a global network of broadband seismometers on Earth and the best limits in this band to date. We also discuss the benefits of a potential Earth-Moon correlation search and compute the time-dependent overlap reduction function required for such an analysis. For this search, we find an upper limit an order of magnitude larger than the Moon-Moon search.
\end{abstract}

\maketitle

\section{Introduction}
\label{sec:Intro}

\begin{figure}[ht!]
\centerline{\hspace*{-0.4cm}\includegraphics[width=1\columnwidth]{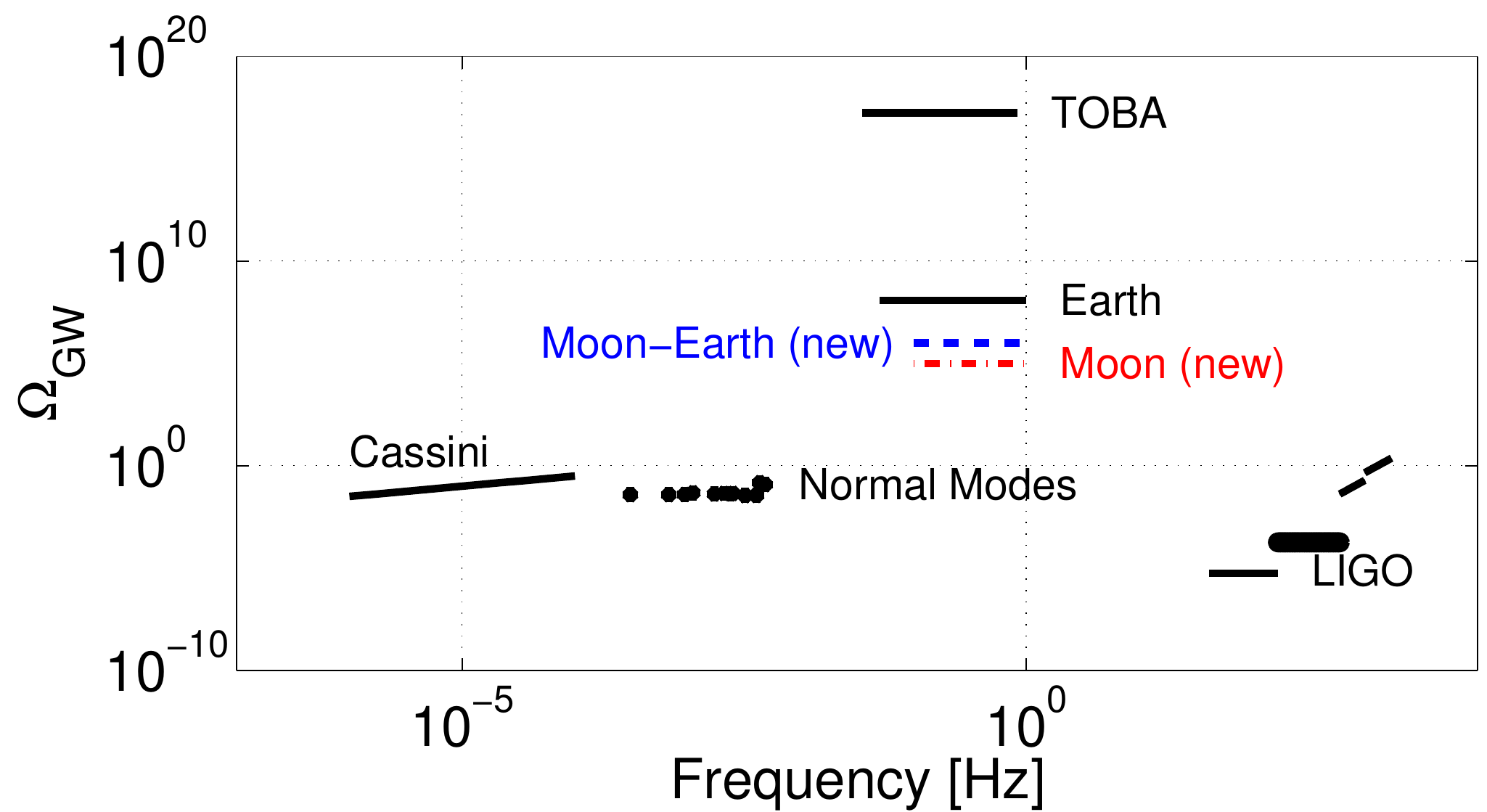}}
\caption{Currently valid upper limits on GW energy density. These limits were set by correlating data from torsion-bar antennas \cite{ShEA2014}, Doppler-tracking measurements of the Cassini spacecraft \cite{ArEA2003}, monitoring Earth's free-surface response with seismometers\cite{CoHa2014}, measuring the normal modes of the Earth \cite{CoHa2014b} and correlating data from the first-generation, large-scale GW detectors LIGO \cite{AaEA2014}.}
\label{fig:bounds}
\end{figure}

The use of astrophysical bodies, such as the Earth, Moon, Sun, and other stars, as detectors of GWs is well-motivated. Siegel and Roth \cite{SiRo2014} recently used high-precision radial velocity data for the Sun to place upperlimits on a stochastic gravitational-wave (GW) background in the millihertz band. We recently used data from a network of modern global broadband seismometers to set upper limits on a stochastic background of GWs in the 0.05-1\,Hz frequency range, which bested previous limits by 9 orders of magnitude \cite{CoHa2014}. These studies complement GW detectors such as LIGO \cite{AbEA2009} and torsion-bar antennas \cite{ShEA2014}. There are currently few dedicated GW experiments in the frequency range $10^{-4}$\,Hz - 10\,Hz. In this band, compact binaries in their inspiral and merger phase are strong possibilities. Although interesting in their own right, these would be a foreground for the potential detection of primordial GWs. Although BICEP2's recent results \cite{AdAi2014} concerning the B-mode polarization of the CMB background was shown to be consistent with dust \cite{AdEA2014}, further work in this area is important as it would provide confirmation of the theory of inflation. AdEA2014Assuming a slow roll inflationary model, this signal would correspond to a GW energy density spectrum $\Omega_{\rm GW} \approx 10^{-15}$ in the 0.1\,Hz to 1\,Hz band. There are a number of GW experiments which could probe this background. Space-based GW detectors will target the frequency band $10^{-4}$\,Hz - 1\,Hz \cite{Vit2014}. There are also concepts for a number of future low-frequency terrestial GW detectors with sensitivity goals better than $10^{-19} / \sqrt{\rm Hz}$ in the 0.1\,Hz to 10\,Hz band \cite{HaEA2013}. The currently valid upper limits are summarized in figure \ref{fig:bounds}.

One of the lessons from the previous study was that a quieter seismic environment would increase the sensitivity of the analysis. For this reason, we analyze Lunar seismic data taken during the Apollo missions. The Moon is the only cosmic body in the Solar system (apart from the Earth), for which seismic data are available. These seismometers were placed on the Moon by the Apollo 12, 14, 15 and 16 missions from 1969 through 1972 and were functional until they were switched off in September 1977. The network was placed on the front center of the Moon in an (approximately) equilateral triangle with 1100-km spacing between stations. Each seismic station consisted of three long-period seismometers aligned orthogonally to measure the three directions of motion. It also included a single-axis short-period seismometer sensitive to vertical motion at higher frequencies. Many important analyses have been performed using this data set, including generation of the first models of the Moon (for a review of the differences between the structures of the Earth, Moon, and Mars, please see \cite{LoMo1993}).

The main sources of ambient noise on the Earth are active tectonics as well as ocean microseisms and atmospheric fluctuations. As the Moon has none of these, the ambient noise background is quite different. Instead, the Moon's seismic noise is predominantly due to tidal forces, thermal stresses, and impacts from asteroids \cite{SeLa2010}. Thermal stresses are due to energy from the Sun, and thus their amplitude is strongly correlated to the lunation period of 29.5 Earth days. They increase after the lunar sunrise and gradually decrease after the sunset. 

More than 12,000 moonquake events have been discovered \cite{Nak2005}. About 7,050 of these events have been positively identified as deep moonquakes, 1,743 are meteoroid impacts, 28 are shallow moonquakes, and there are some other events, such as thermal moonquakes and artificial impacts, and unclassified events. The deep moonquakes are believed to be caused by tidal forces between Earth and the Moon, unlike earthquakes, which are due to tectonic plate movement. The energy loss due to both the deep and shallow moonquake events has been analyzed \cite{GoDa1981}. Deep moonquakes release several orders of magnitude less energy than shallow moonquakes. The moonquakes were analyzed in detail in order to study the shallow seismic velocity structure under the seismometers \cite{CoKo1974,HoLa1980,Nak2011,LaKh2005,SeLa2008}. More recently, analyses of these data suggested a presence of a solid inner and fluid outer core, overlain by a partially molten boundary layer \cite{WeLi2011}. There are proposals to place a seismic array on the Moon to improve on these analyses \cite{Fou2010}. 

In this paper, we report on searches for an isotropic stochastic background using Apollo lunar seismometers from 1976. We perform two such analyses, a Moon-Moon correlation search and an Earth-Moon correlation search. The idea is to use the Moon and Earth as response bodies to GWs. We analyze data from a network of seismometers with near optimal search pipelines. For the Moon-Moon case, we find an upper limit approximately three orders of magnitude smaller than the previous best limits in the frequency range of interest. For the Earth-Moon case, the upper limit is approximately an order of magnitude larger. These are likely to remain the best limits in this frequency band until second-generation torsion bar antennas \cite{IsEA2011}. In section~\ref{sec:formalism}, we outline the search pipeline. We describe the Moon-Moon analysis and results in section~\ref{sec:MoonMoon}. In section~\ref{sec:EarthMoon}, we describe the time-dependent overlap reduction function for an Earth-Moon search and perform a similar analysis. We conclude with a discussion of topics for further study in section \ref{sec:Conclusion}.

\section{Formalism}
\label{sec:formalism}

The response mechanism used here was first described by Dyson \cite{Dys1969}. A rederivation in modern terms was presented in \cite{CoHa2014}. The idea is to measure displacement at the surface due to GWs, given by 
\begin{equation}
\dot\xi_z(\vec r\,,t)\approx-\frac{\beta^2}{\alpha}\vec e_z^\top\cdot h(\vec r\,,t)\cdot \vec e_z,
\label{eq:response}
\end{equation}
where $\vec\xi(\vec r\,,t)$ is ground displacement, $\vec e_z$ the normal vector to the surface, and $h(\vec r\,,t)$ the spatial part of the GW strain tensor (i.~e.~a $3\times 3$ matrix), $\beta$ is the speed of shear waves, and $\alpha$ is the speed of compressional waves. 

\begin{figure*}[ht!]
\includegraphics[width=3.5in]{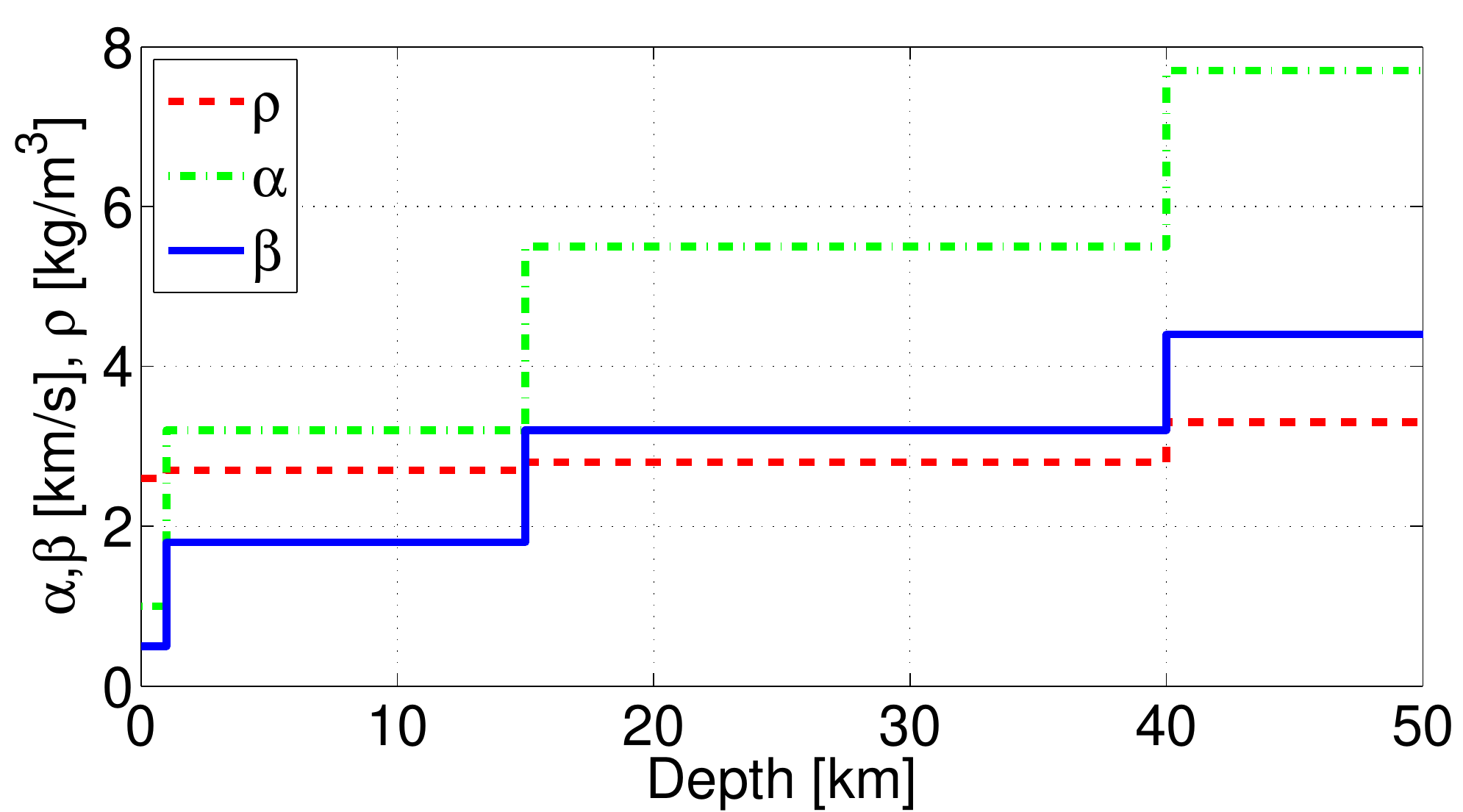}
\includegraphics[width=3.5in]{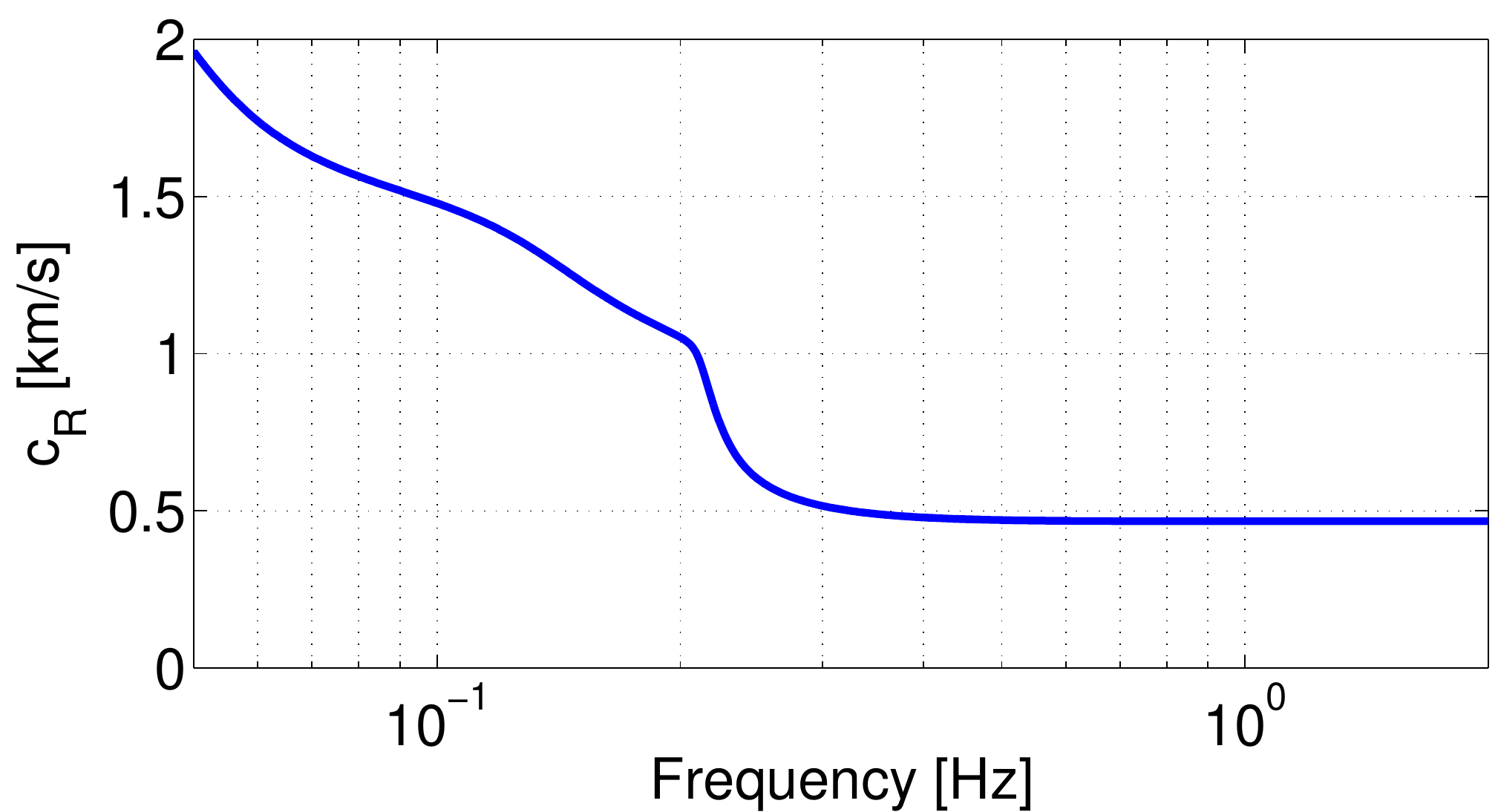}
\caption{On the left is the P-, S-wave velocity and density structure for the whole-Moon model displayed for the near-surface layers (0\,km–-50\,km) \cite{WeLi2011}. On the right is the dispersion curve calculated using Geopsy's gplivemodel for this layered model.}
\label{fig:VpVs}
\end{figure*}
There are two calibration steps required to calibrate raw seismometer data into GW strain. The first is to calibrate raw data of seismometers into ground velocity. Apollo lunar data are sampled in displacement, and so a derivative is first taken to convert to velocity. The second is to calibrate from ground velocity into GW strain. From, Eq.~(\ref{eq:response}), this requires calculation of $\beta^2/\alpha$. To do so, we use a combination of the Poisson's ratio $\nu$ and Rayleigh-wave velocities $c_{\rm R}$ calculated from Weber's Lunar seismic speed model \cite{WeLi2011}. Between 1\,km - 15\,km, $\nu$ assumes values between 0.24 -- 0.27 calculated from the estimates of seismic speeds in each layer. The reason why we do not directly calculate the calibration factor from estimates of $\alpha,\,\beta$ is that these parameters vary with depths, and therefore one needs an effective value of $\beta^2/\alpha$ characterizing the coupling to a surface source (the GW excitation) in a specific frequency range. The corresponding averaging over near surface layers is equivalent to calculating the Rayleigh-wave velocity as a function of frequency. The Rayleigh-wave dispersion curve is obtained using Geopsy's gplivemodel \cite{Wat2013}, which computes dispersion curves from a layered model, and is plotted on the right of Fig.~\ref{fig:VpVs}. From 0.05\,Hz to 0.2\,Hz, the velocities decrease approximately linearly by a factor of 2. From 0.2\,Hz to 0.3\,Hz, the velocities decrease rapidly by another factor of 2. Above 0.3\,Hz, the dispersion is minor. As was argued in \cite{CoHa2014}, only a rough independent estimate of the Poisson's ratio is required to obtain the calibration factor since variations $\Delta\nu$ of the Poisson's ratio are expected to be minor (as seems to be the case as well for the Moon given the range of values obtained from all Moon layers). Expanding the calibration term into linear order of the variation around a reference value of $\nu_0=0.27$, one can estimate the systematic calibration error according to
\beq
\beta^2/\alpha\approx 0.5682c_{\rm R} \cdot(1-1.5377\Delta\nu)
\eeq
Since the Rayleigh-wave speed varies with frequency, the calibration factor is a function of frequency. The gravitational-wave energy density spectrum is defined as
\begin{equation}
\Omega_{\mathrm{GW}} (f) = \frac{1}{\rho_c} \frac{d \rho_c}{d ln f}
\end{equation}
where the critical density of the universe is $\rho_c = 3H_0^2/8 \pi G$. Expressing in terms of the one-sided power spectral density, $S_h (f)$,
\begin{equation}
\Omega_{\mathrm{GW}} (f) = \left(\frac{2 \pi^2}{3 H_0^2} \right) f^3 S_h (f)
\end{equation}
We use a cross$\mbox{-}$correlation method optimized for detecting an isotropic SGWB using pairs of detectors \citep{AlRo1999}.  This method defines a cross$\mbox{-}$correlation estimator:\newline
\begin{equation}
\hat{Y}=\int_{-\infty}^{\infty}df\int_{-\infty}^{\infty}df'\delta_T(f-f'){\tilde s}^{*}_{1}(f){\tilde s}_{2}(f'){\tilde Q}(f')
\end{equation}
and its variance:
\begin{equation}
\sigma^2_Y{\approx}\frac{T}{2}\int_{0}^{\infty}dfP_1(f)P_2(f)|{{\tilde Q}(f)}|^2,
\end{equation}
where $\delta_T(f-f')$ is the finite$\mbox{-}$time approximation to the Dirac delta function, ${\tilde s}_{1}$ and ${\tilde s}_{2}$ are Fourier transforms of time$\mbox{-}$series strain data from two seismometers, $T$ is the coincident observation time, and $P_1$ and $P_2$ are one$\mbox{-}$sided strain power spectral densities from the two seismometers. The signal-to-noise ratio (SNR) can be enhanced by filtering the data \cite{Chr1992,AlRo1999}. The optimal filter spectrum $\tilde Q_{12}(f)$ depends on the overlap-reduction function (ORF) $\gamma_{12}(f)$ \cite{Fla1993}, the noise spectral densities $S_1(f),\,S_2(f)$ of the two seismometers, and also takes into account the relation between the GW spectral density and $\Omega_{\rm GW}$:
\begin{equation}
\tilde Q_{12}(f)=\mathcal{N}\frac{\gamma_{12}(f)}{f^3S_1(f)S_2(f)}
\end{equation}
where $\mathcal{N}$ is a normalization constant \cite{AbEA2012s}. In this form, the filter is optimized for a frequency independent energy density $\Omega_{\rm GW}$. The ORF incorporates the dependence of the optimal filter on the relative positions and alignments of detector pairs. Appendix~\ref{sec:ORF} derives and gives an expression for the ORF in Eq.~\ref{eq:ORF},
where
\beq
\Phi \equiv \frac{2\pi f D}{c}
\eeq
where $D$ is the distance between the seismometers, $f$ is the GW frequency, $c$ the speed of light. The angle $\delta$ denotes the relative orientation of two vertical sensors. In the Moon-Moon case, the distance depends on the location of the seismometers on the Moon
\beq
\begin{split}
D&\equiv 2\sin(\delta/2) R_\circ\\
\sin^2(\delta/2)&=\sin^2(\Delta\lambda/2)+\cos(\lambda_1)\cos(\lambda_2)\sin^2(\Delta\phi/2)
\end{split}
\label{eq:PhiMM}
\eeq
where $R_\circ$ is the Moon's radius, and $\Delta\lambda=\lambda_2-\lambda_1,\,\Delta\phi=\phi_2-\phi_1$ are the differences in latitude and longitude. The distance between a seismometer on Earth and on the Moon is well approximated by the distance between the centers of Earth and Moon, and so that distance is used in the analysis. The change in distance between the Earth and Moon as a function of time is predominantly due to the eccentricity of the Moon's orbit. It is calculated from an implementation of the method described in \cite{Me:1991}. The relative orientation angle $\delta$ needs to be calculated as a function of time for the Moon-Earth analysis as outlined in section \ref{sec:EarthMoon}.

\section{Moon-Moon Correlation}
\label{sec:MoonMoon}
Long-period Apollo PSE ALSEP seismometers were run in two modes of data acquisition. The ``peak'' mode has a peak sensitivity at about 0.5\,Hz. The ``flat-response'' mode is about flat from 0.1-1\,Hz in units of ground displacement. There were four seismometers taking data during this period. Three of the seismometers took coincident broadband data from July 1975 to March 1977. These seismometers form an approximately 1100\,km equilateral triangle. From this period, we took a year of data from 1976. The original data were converted into MiniSEED and supplied by GEOSCOPE. During the conversion process, a constant, nominal sample rate was assumed, when in reality the sample rate varied with time because the timing oscillator on board the central station was not temperature controlled. This generates a small timing error and increases the correlation between stations with a 24 hour periodicity slightly. ALSEP seismometer data was also binned in 54\,s records. Blocking of the entire ALSEP data into 54\,s blocks affected the reference voltage of the analog-to-digital converter through the power supply. This is the likely cause of the small spectral line at about 0.81\,Hz, which is about half the period of a logical record of the ALSEP data (64/106 $\approx$ 0.604 seconds).


For the analysis, we divide the strain time series data into 50\% overlapping 100\,s segments that are Hann$\mbox{-}$windowed. It is found that the seismometers show high seismic correlations due to transient events. These are likely predominantly due to moonquakes and asteroids hitting the Moon.
For this reason, we remove the data surrounding all known moonquakes. This removes about 13\% of the original data and greatly reduces seismic correlation between stations. To minimize remaining correlations, we apply stationarity cuts to the data. We calculate the signal-to-noise ratio of each 100\,s segment and find the signal-to-noise ratio threshold required to bring the point estimate to half of the sigma error bars. We then remove the segments with signal-to-noise ratios that exceed this limit. This removes about 1\% of the data. Ultimately, these vetoes were exclusively related to strong events that contributed to the high-energy tail of the distribution. 

\begin{figure}[t]
 \includegraphics[width=3.5in]{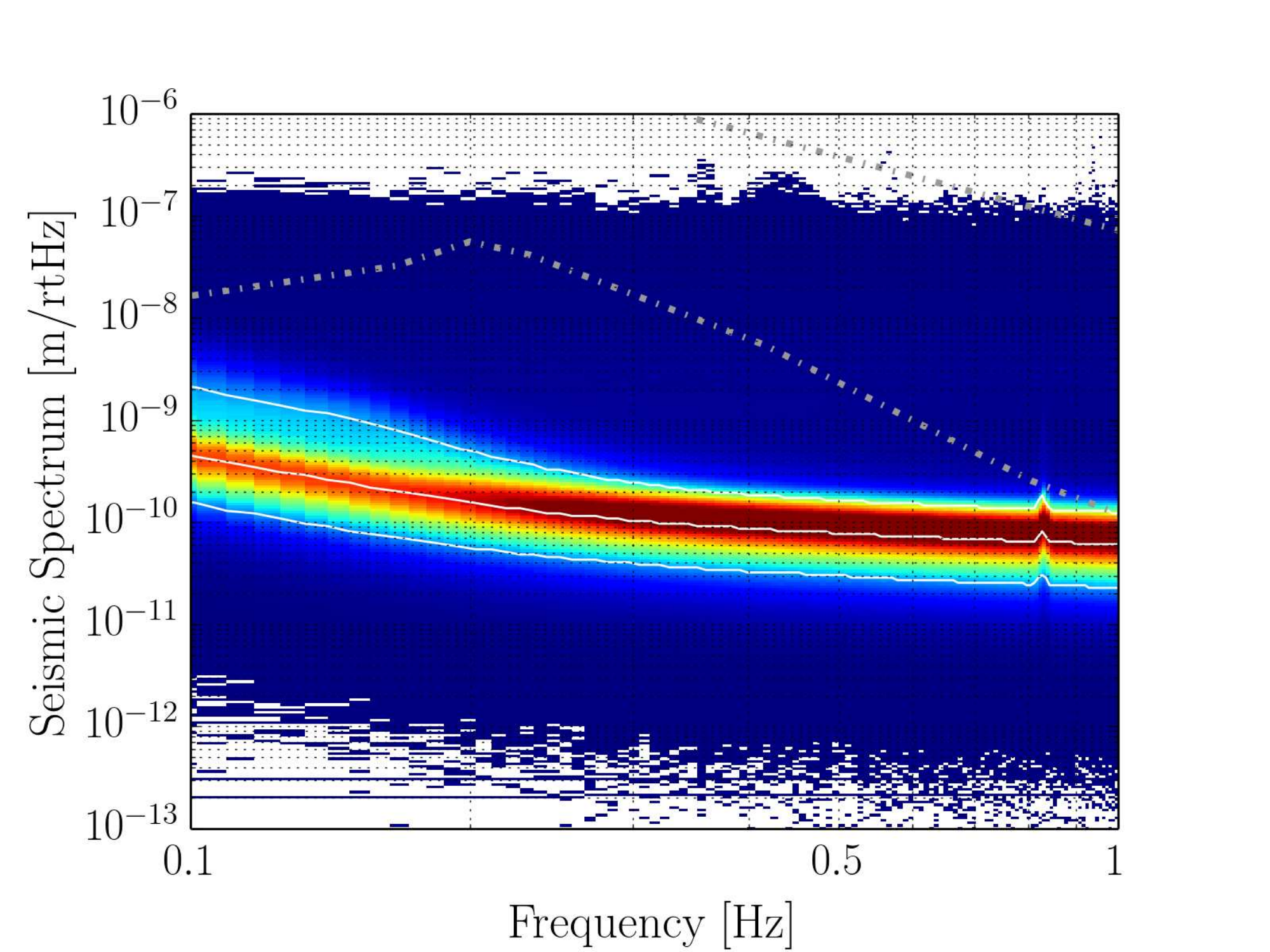}
 \caption{Spectral variation of combined lunar seismic spectra. The dash-dotted lines in gray represent the global new low- and high-noise models \cite{Pet1993}.}
 \label{fig:histoSEIS}
\end{figure}
The vetoed data are excluded from any of the presented results. As a first step, we give a simple characterization of the seismic data in terms of observed seismic spectra. Measuring a seismic spectrum every 128\,s for each seismometer used in the analysis, and combining all these spectra into one histogram, one obtains the result shown in Fig.~\ref{fig:histoSEIS}. This demonstrates the sensitivity of the Lunar seismometers. One can see that the median of the spectra is well below the global new low-noise model \cite{Pet1993}. This low-noise floor of the Earth's ambient noise determined the ultimate noise limit for the Earth analysis. This demonstrates the significant benefit of a lunar search.

To combine the measured $\hat{Y}$ for each of the seismometer pairs, we follow \citep{AlRo1999} and average results from detector pairs weighted by their variances.  The optimal estimator is given by
\begin{equation}
{\hat{Y}_{\rm tot}=\frac{\sum_l\hat{Y}_l\sigma_l^{-2}}{\sum_l\sigma_l^{-2}}}
\end{equation}
where $l$ sums over detector pairs.  The total variance, $\sigma^2_{\rm tot}$, is
\begin{equation}
{\sigma_{\rm tot}^{-2}=\sum_l\sigma_l^{-2}.}
\end{equation}
The resulting combined upper limit of a frequency-independent energy density using the three seismometer pairs, and integrating over frequencies between 0.1\,Hz and 1\,Hz, including calibration errors is
\begin{equation}
\Omega_{\rm GW}<1.2\times 10^{5}
\end{equation}
We assumed a value $H_0=67.8\,\rm km/s/Mpc$ of the Hubble constant \cite{AdEA2013}. Using $S_{\rm GW}(f) = 3H_0^2\Omega_{\rm GW}/(10\pi^2 f^3)$, this translates into a strain sensitivity of about $4.1\times 10^{-15}\,\rm Hz^{-1/2}$ at 0.1\,Hz.
\begin{figure}[ht!]
\centerline{\hspace*{-0.4cm}\includegraphics[width=1\columnwidth]{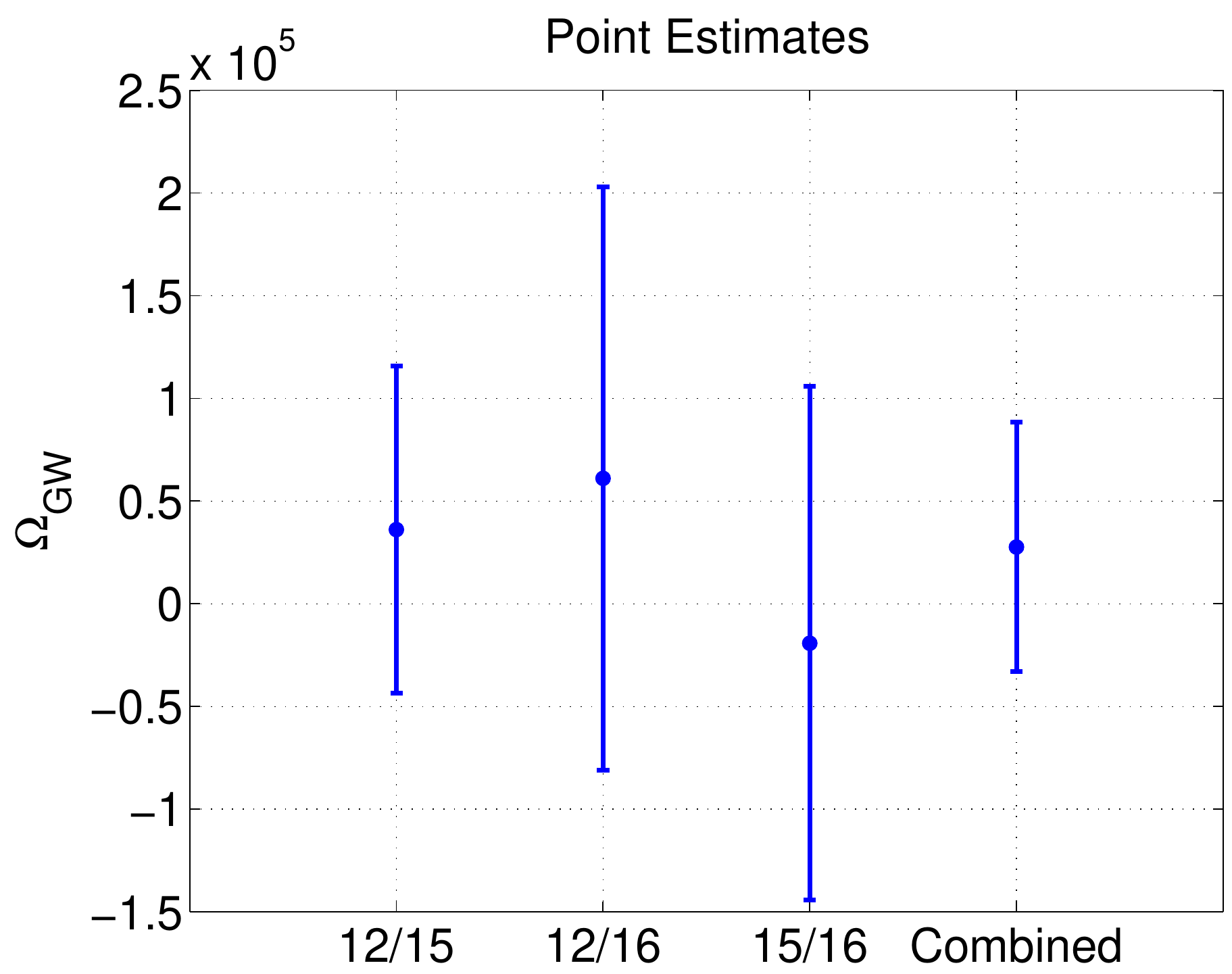}}
\caption{Point estimate and error bars for the three seismometer pairs, as well as for the combination of the three pairs.}
\label{fig:upperlimits}
\end{figure}

\section{Earth-Moon Correlation}
\label{sec:EarthMoon}
In the analyses that have been performed thus far, first on the Earth and now with the Lunar seismometers, the relative orientation of the seismometers analyzed were in a constant orientation. These studies have benefited from the years of seismic data available, both from the vast seismic arrays on Earth as well as the Lunar seismic array. A short-coming of these analyses are the potential for residual seismic correlation due to seismic activity present at both seismometers. 
\begin{figure}[t]
\centerline{\hspace*{-0.4cm}\includegraphics[width=1\columnwidth]{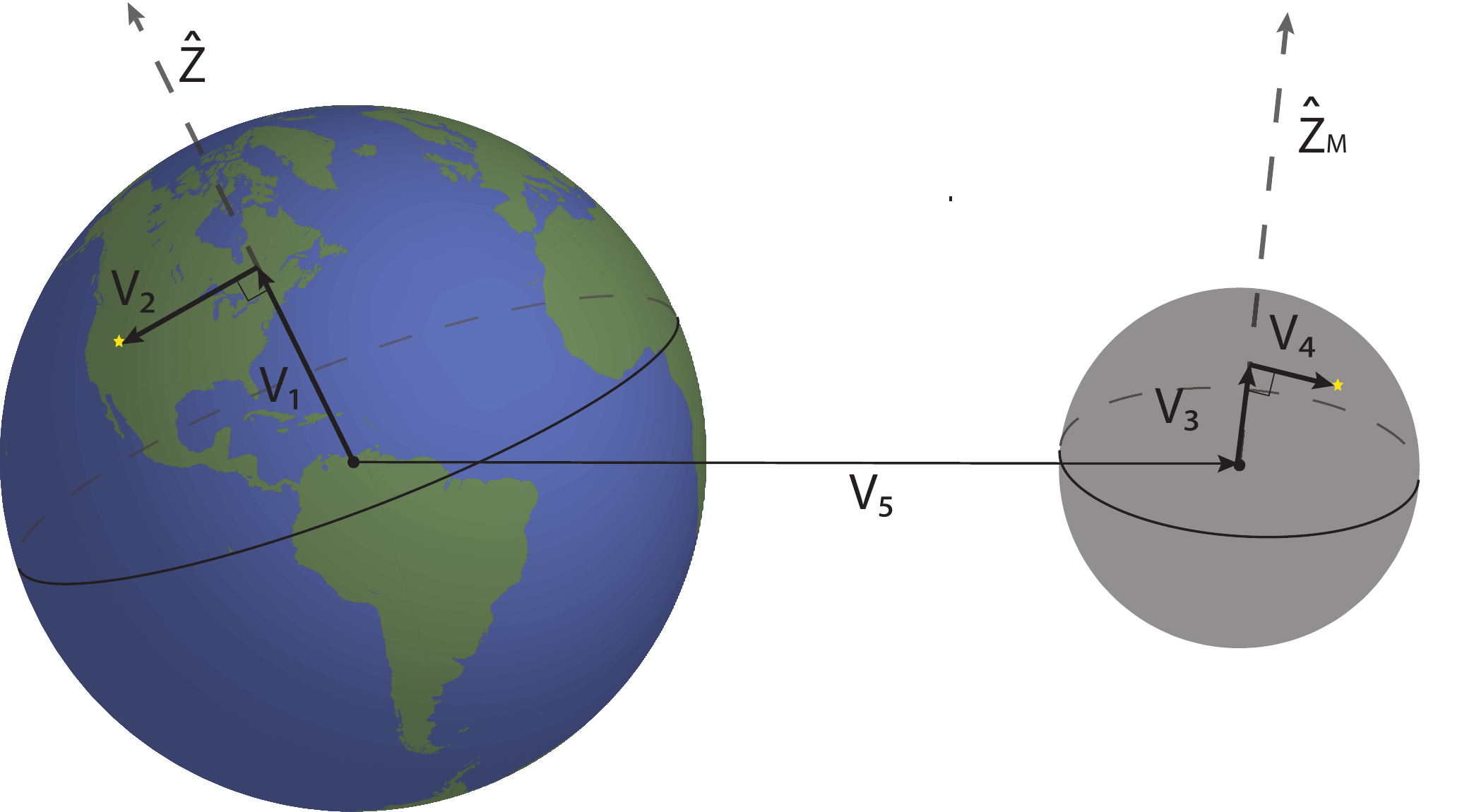}}
\caption{Geometry of the Earth-Moon correlation derivation. The sum of $\vec{v_1}$ and $\vec{v_2}$ correspond to the Earth seismometer, while the sum of $\vec{v_3}$ and $\vec{v_4}$ correspond to the Moon seismometer.}
\label{fig:geometry}
\end{figure}

A potential way around this is to correlate seismometers from the Moon and the Earth. First, we calculate the relative orientation of seismometers on the Earth and the Moon, accounting for the rotation of the bodies. We will then use these orientations to calculate a time-dependent ORF using Eq.~(\ref{eq:ORF}). This calculation will be useful for any situation where detectors have a time-dependent orientation. We begin with the geometry shown in Fig.~\ref{fig:geometry} and define the coordinate system we will work in. A convenient coordinate system to work in is known as the EME2000 system, which is an Earth-centered frame. It defines its epoch to be January 1, 2000, 12 hours Terrestrial Time, which has a Julian Date of 2451545.0. Its $z$-axis is orthogonal to the Earth's mean equatorial plane at this epoch, the $x$-axis aligned with the mean vernal equinox, and the $y$-axis is orthogonal to both, rotated $90^\circ$ about the celestial equator. We denote the unit vectors along these coordinate axes by $\hat{z}$, $\hat{x}$, and $\hat{y}$ respectively. The orientation of the Moon as a function of time is well-known in this system, and therefore provides an ideal coordinate system on which to base these calculations. We make the approximation that the Earth rotation axis has changed negligably between 1970 and 2000, which is reasonable because the Earth's precession has a period of 26,000 years.


We compute each seismometer location as the sum of two orthogonal vectors. For convenience, we will denote three dimensional rotations by $R(\theta,\vec{v})$, where $\theta$ is the angle of rotation and $\vec{v}$ is the vector being rotated around. We denote the longitudes of the Earth and Moon seismometers by $\phi_\mathrm{E}$ and $\phi_\mathrm{M}$ respectively, and the latitudes by $\lambda_\mathrm{E}$ and $\lambda_\mathrm{M}$. We begin with the vectors for the Earth seismometer,  $\vec{v}_1$ and $\vec{v}_2$. The first vector, $\vec{v}_1$, is orthogonal to the Earth's equatorial plane. Because the coordinate system is fixed such that the equator is always in the $x-y$ plane, this vector is a constant in the Z direction with a length of sin($\lambda_\mathrm{E}$). The second vector, $\vec{v}_2$, points from the tip of $\vec{v}_1$ to the seismometer. Because of the Earth's axial rotation, this vector is time-dependent and traces out a circle in the $x-y$ plane, with a radius equal to cos($\lambda_\mathrm{E}$). The vector's angle in this plane can be computed as the sidereal time at the longitude of the seismometer, which corresponds to the seismometer's right ascension in this system. We denote this angle by $\alpha(\phi_\mathrm{E},t)$. Therefore, $\vec{v}_2$ results from a rotation of a vector in the $x$-direction around the $z$-axis by this angle, $R(\mathrm{\alpha(\phi_\mathrm{E},t)},\hat{z})$.

We now perform the same calculation for the Lunar seismometer vectors, $\vec{v}_3$ and $\vec{v}_4$. In this system, the axis orthogonal to the Moon's equator is slightly more complicated than that of the Earth. There are equations describing the right ascension, $\alpha_\mathrm{M}$, and declination, $\delta_\mathrm{M}$, of the Lunar pole as a function of time in the EME2000 coordinate system \cite{ArAH2011}. We denote the vector normal to the Moon's equator, computed from these coordinates using the usual spherical to Cartesian conversion, as $\hat{z}_\mathrm{M}(t)$. $\vec{v}_3$ is then in the direction of $\hat{z}_\mathrm{M}(t)$ with a length sin($\lambda_\mathrm{M}$). To calculate $\vec{v}_4$, we first compute the vector known as the IAU node, which is the vector orthogonal to $\hat{z}_\mathrm{M}$ and the $z$-axis, $\hat{z}_\mathrm{M}(t)\times \hat{z}$. There are also equations describing the angle $W(t)$ between the prime meridian of the Moon and the IAU node vector \cite{ArAH2011}. We use the code in \cite{Ea2013}, which provides $\alpha_\mathrm{M}$, $\delta_\mathrm{M}$, and W, to compute these quantities. To account for the difference between the prime meridian and the longitude, we add the Lunar seismometer's longitude to $W(t)$. $\vec{v}_4$ is then computed by rotating the IAU node by this angle, $R(W(t)+\phi_\mathrm{M},\hat{z}_\mathrm{M}(t))$. It has a length of $\cos(\lambda_\mathrm{M})$.

We can now succinctly summarize the computation of the four vectors in equation form:
\beq
\begin{split}
\vec{v}_1 &= \sin(\lambda_\mathrm{E})\hat{z}\\
\vec{v}_2 &= \cos(\lambda_\mathrm{E})R(\alpha(\phi_\mathrm{E},t),\hat{z}) \cdot \hat{x} \\
\vec{v}_3 &= \sin(\lambda_\mathrm{M})\hat{z}_\mathrm{M}(t) \\
\vec{v}_4 &= \cos(\lambda_\mathrm{M})R(W(t)+\phi_\mathrm{M},\hat{z}_\mathrm{M}(t)) \cdot \frac{\hat{z}_\mathrm{M}(t) \times \hat{z}}{\|\hat{z}_\mathrm{M}(t) \times \hat{z}\|}
\end{split}
\label{eq:vectors}
\eeq
where the Moon rotation axis vector is given in terms of the time-dependent right-ascension and declination angles according to
\begin{equation}
\hat{Z}_\mathrm{M} = [\cos(\alpha_\mathrm{M})\cos(\delta_\mathrm{M}),\sin(\alpha_\mathrm{M}) \cos(\delta_\mathrm{M}),\sin(\delta_\mathrm{M})]
\end{equation}
We now have the required quantities to calculate the angle $\delta$ in Eq.~(\ref{eq:ORF}) between the two seismometers. This corresponds to finding the angle between $\vec{v}_1+\vec{v}_2$ and $\vec{v}_3+\vec{v}_4$. The final ingredient for computing the ORF is the distance between the two seismometers. $\vec{v}_5$ shows the vector connecting the center of the Earth to the center of the Moon. It is assumed that the distance between a seismometer on Earth and on the Moon is well approximated by the distance between the centers of Earth and Moon. Expanding the ORF in small changes of the seismometers' distance, one finds that it depends linearly on the change in distance in general. Since the relative change in distance is about 0.03, the relative change in ORF is of similar order. This lies well below other modelling errors, such as the seismic-speed dependent calibration term in Eq.~(\ref{eq:response}), and therefore we can conclude that the approximation does not introduce significant systematic errors.

\begin{figure}[t]
\centerline{\hspace*{-0.4cm}\includegraphics[width=1\columnwidth]{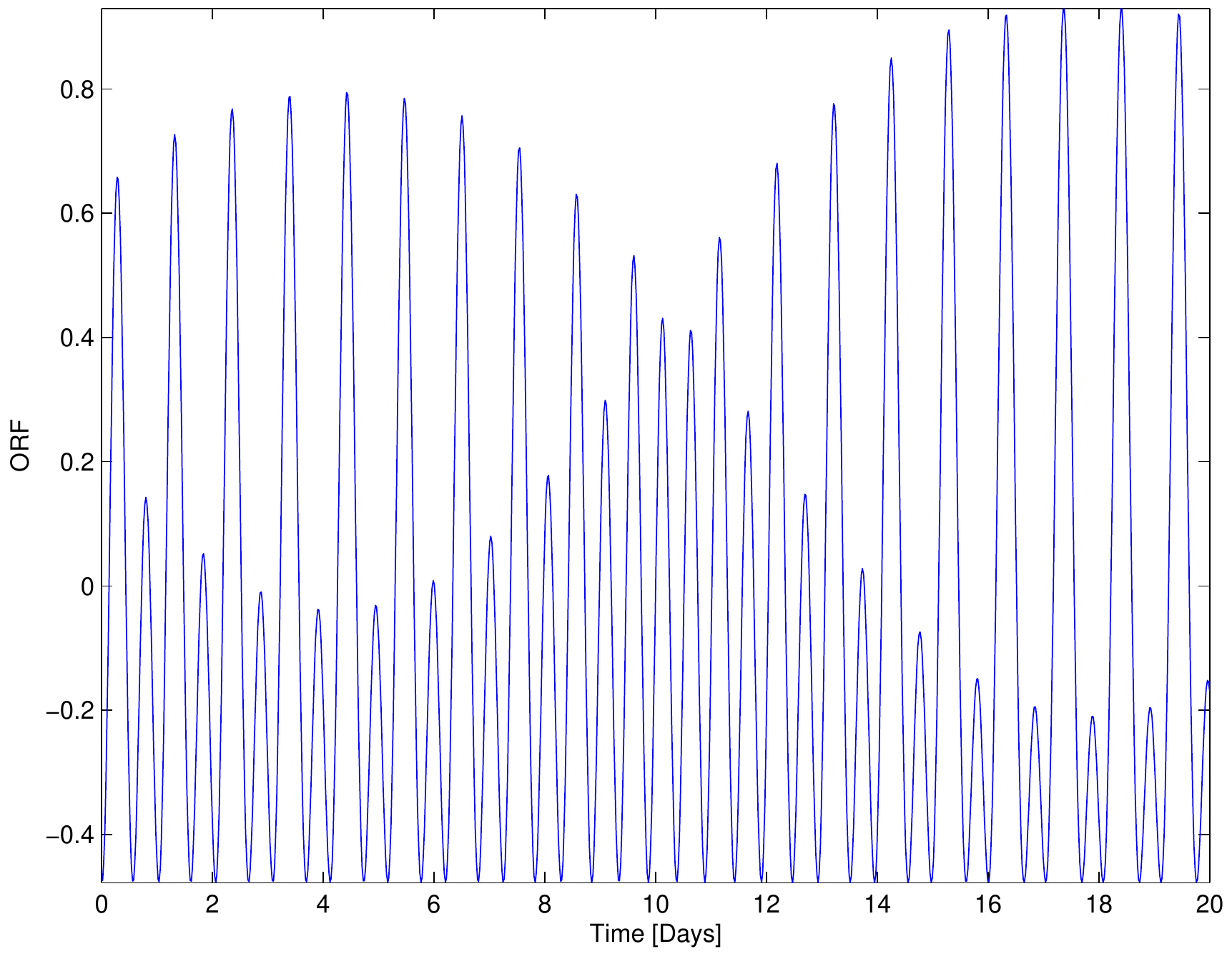}}
\caption{Overlap reduction function between example Earth and Moon seismometers as a function of time at 0.1\,Hz. This is computed for the S12 Lunar station and Albuquerque, New Mexico, USA (ANMO) seismic stations at the beginning of 1976. It displays a daily variation due to the rotation of the earth, as well as a monthly cycle, due to the libration of the Moon.}
\label{fig:angles}
\end{figure}

Figure~\ref{fig:angles} shows the ORF between example Earth and Moon seismometers as a function of time at 0.1\,Hz. The ORF displays both a daily cycle, due to the rotation of the Earth, as well as a monthly cycle, due to the Moon's libration, which has a 28 day period. Due to tidal locking, the Moon has only one face pointing towards the Earth at all times. The face pointing towards the Earth oscillates slightly with time, and this effect is called libration. Slightly more than half of the Moon's surface can be seen from Earth. Libration in longitude results from the eccentricity of the Moon's orbit around Earth, while libration in latitude results from the inclination between the Moon's axis of rotation and the normal of its orbital plane.

We perform an analysis as described above using the Lunar seismic array and seismometers on Earth. The Earth seismometers used are Geotech KS-36000 Borehole seismometers located in Albuquerque, New Mexico, USA (ANMO), Guam, Marianas Islands (GUMO), and Mashhad, Iran (MAIO). To calibrate these data, we convert from displacement to velocity by first taking a derivative. Following \cite{CoHa2014}, we use a global phase velocity map of $c_{\rm R}$ to calculate $\beta^2/\alpha$ for the Earth seismometers \cite{Ek2011}. We perform the analysis on the coincident data from 1976 for all nine possible pairs, which yields the following upper limits for all pairs and the single best pair:
\begin{equation}
\Omega^{\rm tot}_{\rm GW}<1.1\times 10^{6},\,\Omega^{\rm sgl}_{\rm GW}<1.5\times 10^{6}
\end{equation}
Although this is an order of magnitude higher than that obtained for the Moon-Moon analysis, it is in some sense a more robust upper limit since correlation between seismometers can only exist due to a GW signal.

\section{Conclusion}
\label{sec:Conclusion}

The results presented in this paper using the Apollo Lunar seismometers are likely the best upper limits that can currently be achieved with seismometers in the frequency range 0.1\,Hz -- 1\,Hz. As the Moon has the lowest ambient seismic noise currently measured, future improvements on this upper limit using seismic measurements should not be expected. For this reason, these constraints are likely to remain the best in this frequency band until second-generation torsion bar antennas \cite{IsEA2011}. Unlike indirect limits that exist on cosmological backgrounds, our method directly constrains the astrophysical and cosmological components of the stochastic GW background. This complements the direct upper limits in other frequency bands and the integrated, indirect upper limits in the same frequency regime.

There are proposals to create a network of lunar geophysical stations, such as SELENE2 \cite{TaSh2008}, and the development of a network of broadband seismometers on the surface of the Moon \cite{YaGa2011}. These are in conjunction with space missions being planned to create a lunar station as well as to do fundamental science, including constructing a theory of the formation of the Earth, and its initial state and evolution. The use of an array of modern seismometers placed in an anti-podal network array would likely improve the result, although not likely more than an order of magnitude. 

\begin{acknowledgements}
MC was supported by the National Science Foundation Graduate Research Fellowship
Program, under NSF grant number DGE 1144152. The stochastic GW search has been carried out using the MatApps software available at \url{https://www.lsc-group.phys.uwm.edu/daswg/projects/matapps.html}. Seismic data were downloaded from servers of GEOSCOPE (\url{http://geoscope.ipgp.fr/index.php/en/}). We greatly appreciate the helpful comments about Apollo seismic data from Yosio Nakamura and the overlap reduction function from Joseph Romano. We would like to thank Valerie Fox for help in designing Fig.~\ref{fig:geometry}.
\end{acknowledgements}

\appendix

\section{Calculation of overlap-reduction function}
\label{sec:ORF}
The overlap reduction function (ORF) can be calculated using the results published in \cite{AlRo1999}. The main calculation is summarized here also to emphasize that the solution presented in \cite{CoHa2014} is not fully general as implied in their paper. It is an accurate approximation to the fully general ORF in the case of antipodal seismometer pairs. In the following, we outline the calculation of the fully general solution. The goal is to calculate the correlation of GW signals between two seismometers measured according to Eq.~(\ref{eq:response}). The method is to express this correlation for GWs incident from a specific direction and with specific polarization, and then to average the expression over propagation directions and wave polarizations. Without applying further weights, this yields the overlap-reduction function relevant to an isotropic, stochastic GW background. We will provide an explicit solution for the case of two seismometers on the Moon (or Earth). This however can not be applied to the Moon-Earth analysis, since the explicit solution is only for the specific geometry of two seismometers on the same sphere. Generalized expressions of the results in \cite{AlRo1999} have to be employed for the Moon-Earth correlations. 

We start with the Moon-Moon correlation. In a Moon-centered spherical coordinate system, the spatial components of the strain tensor for the two polarizations of a GW can be written as
\beq
\begin{split}
\mathbf{h}_+(\vec r,t) &= (\hat\theta\otimes\hat\theta-\hat\phi\otimes\hat\phi)h_+{\rm e}^{\irm (\omega t-\vec r\cdot\vec k)}\\
&= \mathbf{e}_+h_+{\rm e}^{\irm (\omega t-\vec r\cdot\vec k)},\\
\mathbf{h}_\times(\vec r,t) &= (\hat\theta\otimes\hat\phi+\hat\phi\otimes\hat\theta)h_\times{\rm e}^{\irm (\omega t-\vec r\cdot\vec k)}\\
&= \mathbf{e}_\times h_\times{\rm e}^{\irm (\omega t-\vec r\cdot\vec k)}
\end{split}
\eeq
where $\hat\theta,\,\hat\phi$ are the unit normal vectors of the two angular coordinate axes. There are various ways to average over polarizations. One can either introduce a polarization angle that mixes these two elementary polarizations and average over it, or one can follow the usual recipe to calculate the correlation between the two detector signals for each polarization, and then add the terms (different polarizations are mutually uncorrelated). Following the last method, the ORF for a seismometer pair can be written as
\beq
\begin{split}
&\gamma_{12}(f)=\\
&\;\frac{15}{32\pi}\sum\limits_{A=+,\times}\int\drm\Omega_k (\hat z_1^\top\cdot\mathbf{e}_A\cdot\hat z_1)(\hat z_2^\top\cdot\mathbf{e}_A\cdot\hat z_2){\rm e}^{\irm 2\pi f\hat k\cdot\Delta\vec r/c}
\end{split}
\eeq
The normalization factor is chosen such that the ORF for two collocated seismometers is equal to 1. The unit vectors $\hat z_1,\,\hat z_2$ denote the surface normals at the two seismometer locations, and $\Delta\vec r$ is the separation vector between the two seismometers. The simplest way to calculate the integrals is to choose $\Delta\vec r$ as the $z$-axis of the spherical coordinate system so that 
\beq
\hat k\cdot\Delta\vec r=|\Delta\vec r\,|\cos(\theta)
\eeq
The integrals are easy to carry out now, and the remaining problem is to transform the result into an elegant form in terms of the angle $\delta$ subtended by the great circle connecting the two seismometers. Substituting Moon coordinates for the two unit vectors $\hat z_1,\,\hat z_2$, one has to keep in mind that the $z$-axis was chosen to lie in the direction of the separation line. Therefore, to derive the final result in an elegant way one needs to choose Moon coordinates of the two seismometers such that this condition is fulfilled. This means that the two longitudes have to be the same, and the two latitudes have opposite sign. Inserting these angles into the Moon-coordinate expressions of $\hat z_1,\,\hat z_2$, and with $\delta$ being the difference between the two latitudes, one directly obtains the equation
\beq
\begin{split}
\gamma_{12}(f)=\frac{15}{64\Phi^2}\bigg(&8\Phi^2\cos^4(\delta/2)j_0(\Phi) \\ &+8\Phi\cos^2(\delta/2)(3\cos(\delta)-5)j_1(\Phi)\\
&+(41-20\cos(\delta)+3\cos(2\delta))j_2(\Phi)\bigg)
\end{split}
\eeq
It can be verified that for $\delta\approx\pi$, i.~e.~near antipodal seismometers, the ORF is well approximated by the expression given in \cite{CoHa2014}, and the result here is identical to the expression one obtains applying the equations of \cite{AlRo1999} to a seismometer pair.
\begin{figure}[t]
\centerline{\includegraphics[width=0.9\columnwidth]{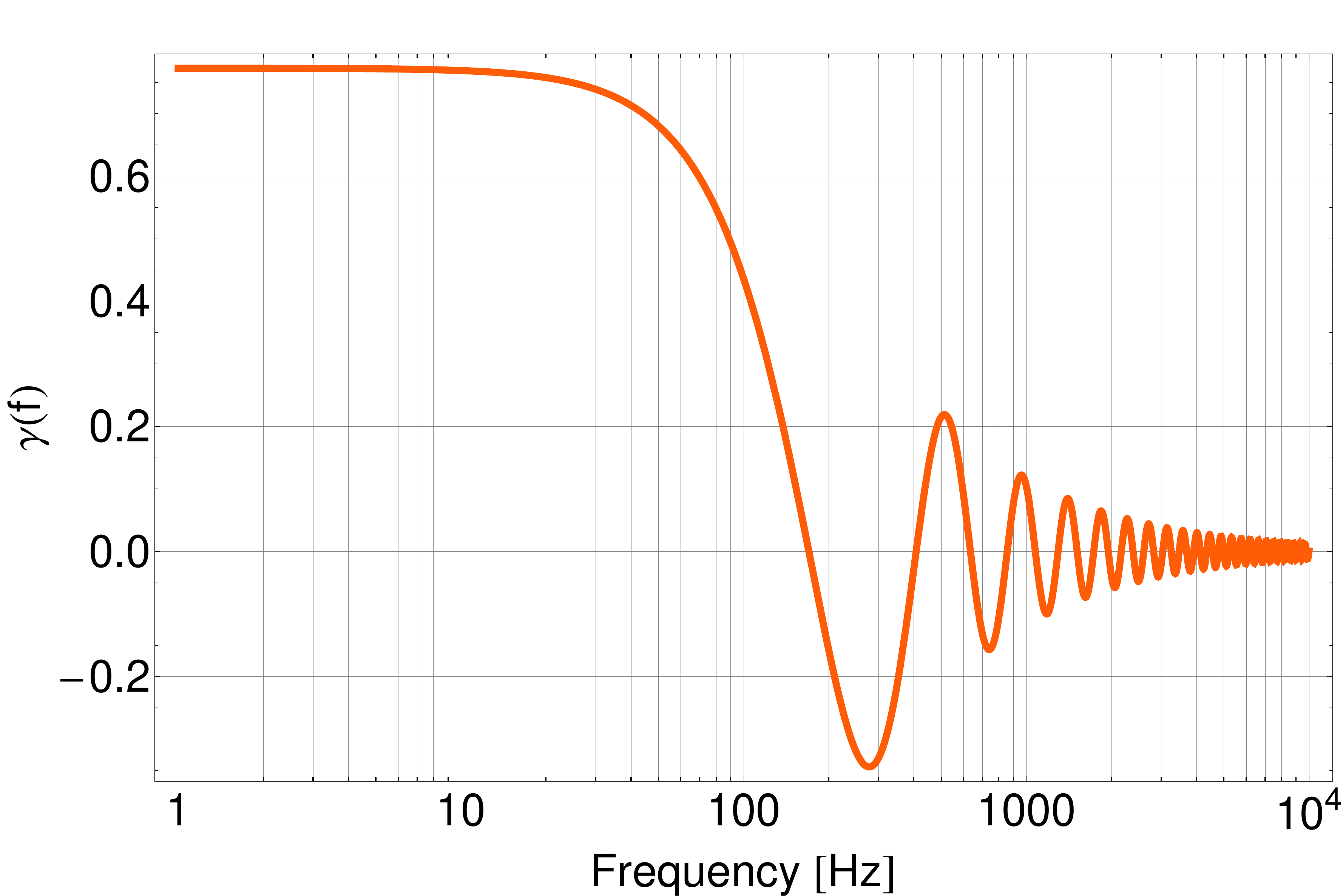}}
\caption{Overlap reduction function between two seismometers on the Moon separated by 0.4\,rad. }
\label{fig:moonorf}
\end{figure}
An example of the ORF is plotted in Fig.~\ref{fig:moonorf} for two Moon seismometers separated by $\delta=0.4\,$rad. Due to the small size of the Moon with respect to the length of a GW, the correlation between the two seismometers is constantly high up to about 10\,Hz. Above 10\,Hz, we can see the typical oscillations produced by the spherical Bessel functions. 

An explicit expression of the correlation between Moon and Earth seismometers will not be given here, but instead, we will use a generalized form of Eq.~(3.42) in \cite{AlRo1999}. Their result can be expressed in terms of the surface normal vectors $\vec e_z^{\,1},\,\vec e_z^{\,2}$ at the two seismometers, and the unit separation vector $\hat s\equiv\Delta\vec r/\Delta r$. In this case, the ORF can be written in terms of the detector response matrices $\mathbf{d}_1\equiv\vec e_z^{\,1}\otimes\vec e_z^{\,1}$, $\mathbf{d}_2\equiv\vec e_z^{\,2}\otimes\vec e_z^{\,2}$
\beq
\begin{split}
\gamma_{12}&(f)=\\
&\frac{3}{4}\big(A(\Phi){\rm Tr}(\mathbf{d}_1){\rm Tr}(\mathbf{d}_2)+2B(\Phi){\rm Tr}(\mathbf{d}_1\mathbf{d}_2)\\
&+C(\Phi)((\hat s^\top\mathbf{d}_1\hat s){\rm Tr}(\mathbf{d}_2)+{\rm Tr}(\mathbf{d}_1)(\hat s^\top\mathbf{d}_2\hat s))\\
&+4D(\Phi)s^\top(\mathbf{d}_1\mathbf{d}_2)\hat s+E(\Phi)(\hat s^\top\mathbf{d}_1\hat s)(\hat s^\top\mathbf{d}_2)\hat s)\big)
\end{split}
\label{eq:ORF}
\eeq
The original result given in \cite{AlRo1999} is obtained for trace-free detector matrices $\mathbf{d}_1,\,\mathbf{d}_2$. As usual, the ORF is normalized to be equal to 1 for collocated seismometers. The coefficients $A(\Phi),\,B(\Phi),\,C(\Phi),\,D(\Phi),\,E(\Phi)$ are linear combinations of the three spherical Bessel functions $j_0(\Phi),\,j_1(\Phi),\,j_2(\Phi)$. This expression can be applied to arbitrary seismometer locations and orientations, and for the Earth-Moon analysis, depending on the chosen coordinate frame, potentially all three vectors $\hat s,\,\vec e_z^{\,1},\,\vec e_z^{\,2}$ are functions of time.

\bibliographystyle{unsrt}
\bibliography{references}

\end{document}